\documentclass{ws-procs975x65}

\begin{document}

\title{Hybrid neutron stars within the \\ Nambu-Jona-Lasinio model and confinement}

\author{M. Baldo, G. F. Burgio, P. Castorina, S. Plumari, and D. Zappala'}

\address{Dipartimento di Fisica e Astronomia,Universita' di Catania and INFN,\\ 
Sezione di Catania, Via S. Sofia 64, I-95123 Catania, Italy}

\begin{abstract}
Recently, it has been shown that the standard Nambu-Jona-Lasinio (NJL) model is not 
able to reproduce the correct QCD behavior of the gap equation at 
large density, and therefore a different cutoff procedure at large 
momenta has ben proposed. We found that, even with this density 
dependent cutoff procedure, the pure quark phase in neutron stars (NS) interiors is unstable, 
and we argue that this could be related 
to the lack of confinement in the original NJL model. 
\end{abstract}


\bodymatter

\section{NJL AT LARGE DENSITY}\label{aba:sec1}
In the interior of astrophysical compact objects, like NS, 
nuclear matter is expected to reach a density which is several times nuclear 
saturation density. In these conditions, calculations only based on nucleonic 
degrees of freedom becomes highly questionable and a phase transition to quark 
matter becomes possible. The quark matter equation of state (EoS) derived from 
standard NJL model is soft enough to render NS unstable at the onset of 
the deconfined phase, and no pure quark matter can be present in its interior. 
Though the NJL model reproduces correctly the phenomenological low energy data on 
hadron properties, it is not able to reproduce the correct behavior of the
solution of QCD gap equation  at large density, as pointed out in Ref. \cite{casa}. 
In order to clarify this point, we have studied a modified NJL model with a density 
dependent momentum cutoff, which preserves the low energy properties of the theory.
According to this procedure, a $\mu$ dependent cut-off $\Lambda(\mu)$ is 
introduced, which implies a $\mu$ dependent coupling constant $g(\mu)$. 
Since there is no compelling restriction to a specific functional form 
of $\Lambda(\mu)$, we choose some smooth monotonically increasing functions of $\mu$.
For details, the reader is referred to Ref. \cite{baldo}.
Considering different slopes in the $\mu$ dependence of the cut-off
is an important issue, because this implies different growth
behaviors of the pressure, as a function of the baryon chemical 
potential $\mu_B$, which is crucial to determine the transition point to quark matter.
By assuming a first order phase transition, 
as suggested by the indications coming from lattice calculations \cite{fodor},
we have adopted for the hadronic phase a nucleonic equation of state obtained 
within the Brueckner-Bethe-Goldstone (BBG) approach \cite{col}, 
while for the quark phase, we have used the standard 
parametrization of the NJL model \cite{bubtes} and the above prescriptions.
\begin{figure*}[t]
\centering
\includegraphics[width=4.8 truecm,angle=270]{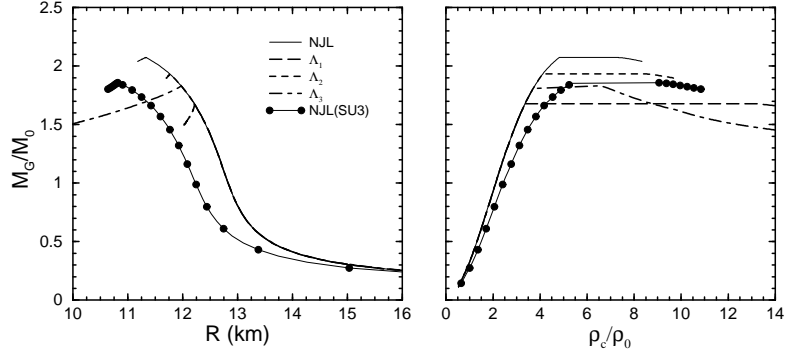}
\caption{The gravitational mass (in units of the solar mass
$M_\odot=2\times 10^{33}g$) is plotted as function of the radius (left panel)
and the central density (right panel), for different choices of the cut-off
behavior.}
\end{figure*}
In Fig. 1 we report the neutron star masses as a function of the
radius (left panel), and of the central density (right panel) for
the different choices of the density dependent cut-off discussed
above. For comparison also the results for the standard NJL model
(three flavor) of Ref. \cite{bub} are reported (full circles). 
The plateau in the mass-central density plane is a consequence 
of the Maxwell construction. In all cases one can see that at the 
maximum mass the plot is characterized by a cusp, which corresponds 
to the instability mentioned above. 
As discussed in Ref. \cite{baldo} the origin of the NS instability could 
be related to the missing quark confinement in the model.
In fact, when chiral symmetry is restored, the NJL model behaves like 
the MIT model with a bag constant $B_{NJL}\simeq 140 \; MeV/fm^{3}$. 
Therefore, if one adds by hand a confining potential which is switched 
off at the chiral phase transition \cite{thom}, the instability could be removed 
since the effective bag constant would be correspondingly reduced.



\begin{thebibliography}{99}
\bibitem{baldo} M. Baldo, G.F. Burgio, P. Castorina, S. Plumari and 
D. Zappala', Phys. Rev. C {\bf 75}, 035804 (2007).
\bibitem{casa} R. Casalbuoni, R. Gatto, G. Nardulli and M. Ruggieri, 
Phys. Rev. D {\bf 68}, 034024 (2003).
\bibitem{fodor} Z. Fodor and S. D. Katz, JHEP {\bf 0404}, 050 (2004).
\bibitem{col} C. Maieron, M. Baldo, G. F. Burgio, and H.-J. Schulze, 
Phys. Rev. D {\bf 70}, 043010 (2004), and references therein.
\bibitem{bubtes} M. Buballa, Phys. Rep. {\bf 407}, 205 (2005).
\bibitem{bub} M. Baldo, M. Buballa, G. F. Burgio, F. Neumann, M. Oertel, and
H.-J. Schulze, Phys. Lett. B {\bf 562}, 153 (2003), and references therein.
\bibitem{thom} S. Lawley, W. Bentz and A.W. Thomas, J. Phys. G {\bf 32}, 667
(2006).
\vfill\eject
\end{thebibliography}
\end{document}